\documentclass[12pt]{article}
\pdfoutput=1
\usepackage{bm,amsmath,amssymb,graphicx}

\begin{document}
\pagenumbering{arabic}
\begin{titlepage}

\title{Late cosmology in massive conformal gravity}

\author{F. F. Faria$\,^{*}$ \\
Centro de Ci\^encias da Natureza, \\
Universidade Estadual do Piau\'i, \\ 
64002-150 Teresina, PI, Brazil}

\date{}
\maketitle

\begin{abstract}
In this paper we find the cosmological solutions of the massive conformal 
gravity field equations in the presence of matter fields. In particular, 
we show that the solution of negative curvature is in good agreement with 
the late universe. 
\end{abstract}

\thispagestyle{empty}
\vfill
\noindent PACS numbers: 04.50.Kd, 98.80.-k \par
\bigskip
\noindent * felfrafar@hotmail.com \par
\end{titlepage}
\newpage


\section{Introduction}
\label{sec1}


The massive conformal gravity (MCG) is a conformally invariant theory of 
gravity in which the gravitational action is the sum of the Weyl action with 
the Einstein-Hilbert action conformally coupled to a scalar field 
\cite{Faria1}. At the classical level, it has been shown that the theory is 
free of the vDVZ discontinuity \cite{Faria2} and can reproduce the orbit of 
binaries by the emission of gravitational waves \cite{Faria3}. Furthermore, 
MCG is a renormalizable and unitary quantum theory of gravity 
\cite{Faria4,Faria5,Faria6}. 

Despite the promising results of MCG obtained so far, it is very important to 
test the theory with cosmological observations. The most accepted cosmological 
model to explain the dynamics of the universe is the 
$\Lambda$CDM model. However, this model suffers from important problems such 
as the cosmological constant problem \cite{Weinberg,Carroll} and the 
Hubble tension between the early and late universe observational data 
\cite{Planck,Riess}. Here we want to see if MCG explains the late universe 
without the cosmological constant problem. 

This paper is organized as follows. In Sec. \ref{sec2} we describe the MCG 
cosmological field equations. In Sec. \ref{sec3} we find the MCG 
cosmological solutions. In Sec. \ref{sec4} we compare the cosmological 
solutions of MCG with cosmological observations. Finally, in Sec. \ref{sec5} 
we present our conclusions.


\section{Cosmological field equations}
\label{sec2}


The total MCG action is given by\footnote{This action is obtained from the 
action of Ref. \cite{Faria2} by rescaling $\varphi \rightarrow 
\left(\sqrt{32\pi G/3}\right)\varphi$ and considering 
 $m = \sqrt{3/64\pi G\alpha}$.} \cite{Faria2}
\begin{equation}
S = \int{d^{4}x} \, \sqrt{-g}\bigg[ \varphi^{2}R 
+ 6 \partial^{\mu}\varphi\partial_{\mu}\varphi 
- \frac{1}{2\alpha^2} C^{\alpha\beta\mu\nu}C_{\alpha\beta\mu\nu} \bigg] 
+ \int{d^{4}x\mathcal{L}_{m}},
\label{1}
\end{equation}
where $\varphi$ is a scalar field called 
dilaton, $\alpha$ is a dimensionless constant, 
\begin{equation}
C^{\alpha\beta\mu\nu}C_{\alpha\beta\mu\nu} = R^{\alpha\beta\mu\nu}
R_{\alpha\beta\mu\nu} - 4R^{\mu\nu}R_{\mu\nu} + R^2 
+ 2\left(R^{\mu\nu}R_{\mu\nu} - \frac{1}{3}R^{2}\right)
\label{2}
\end{equation}
is the Weyl tensor squared,
$R^{\alpha}\,\!\!_{\mu\beta\nu} 
= \partial_{\beta}\Gamma^{\alpha}_{\mu\nu} + \cdots$ is the Riemann tensor, 
$R_{\mu\nu} = R^{\alpha}\,\!\!_{\mu\alpha\nu}$ is the Ricci tensor, 
$R = g^{\mu\nu}R_{\mu\nu}$ is the scalar curvature, and 
$\mathcal{L}_{m} = \mathcal{L}_{m}(g_{\mu\nu},\Psi)$ is the Lagrangian 
density of the matter field $\Psi$. It is worth noting that the action 
(\ref{1}) is invariant under the conformal transformations
\begin{equation}
\tilde{g}_{\mu\nu}=e^{2\theta(x)}\,g_{\mu\nu},
\ \ \ \ \
\tilde{\varphi}=e^{-\theta(x)} \varphi,
\ \ \ \ \
\tilde{\mathcal{L}}_{m} = \mathcal{L}_{m},
\label{3}
\end{equation}
where $\theta(x)$ is an arbitrary function of the spacetime coordinates. 

The variation of (\ref{1}) with respect to $g^{\mu\nu}$ and $\varphi$ gives 
the MCG field equations
\begin{equation}
\varphi^{2}G_{\mu\nu} +  6 \partial_{\mu}\varphi\partial_{\nu}\varphi 
- 3g_{\mu\nu}\partial^{\rho}\varphi\partial_{\rho}\varphi + g_{\mu\nu} 
\nabla^{\rho}\nabla_{\rho} \varphi^{2} 
- \nabla_{\mu}\nabla_{\nu} \varphi^{2}  - \alpha^{-2} W_{\mu\nu} 
= \frac{1}{2}T_{\mu\nu},
\label{4}
\end{equation}
\begin{equation}
\left(\nabla^{\mu}\nabla_{\mu} - \frac{1}{6}R \right) \varphi = 0,
\label{5}
\end{equation}
where
\begin{equation}
W_{\mu\nu} = \nabla^{\alpha}\nabla^{\beta}C_{\mu\alpha\nu\beta} 
-\frac{1}{2} R^{\alpha\beta}C_{\mu\alpha\nu\beta} 
\label{6}
\end{equation}
is the Bach tensor,
\begin{equation}
G_{\mu\nu} = R_{\mu\nu} - \frac{1}{2}g_{\mu\nu}R
\label{7}
\end{equation}
is the Einstein tensor,
\begin{equation}
\nabla^{\rho}\nabla_{\rho} \varphi = 
\frac{1}{\sqrt{-g}}\partial^{\rho}\left( \sqrt{-g} \partial_{\rho}
\varphi \right)
\label{8}
\end{equation} 
is the generally covariant d'Alembertian for a scalar field, and
\begin{equation}
T_{\mu\nu} = \frac{2}{\sqrt{-g}} \frac{\delta \mathcal{L}_{m}}
{\delta g^{\mu\nu}}
\label{9}
\end{equation}
is the matter energy-momentum tensor.

At scales below the Planck scale, the dilaton field acquires a spontaneously 
broken constant vacuum expectation value $\varphi_{0}$ \cite{Matsuo}. In 
this case, the field equations (\ref{4}) and (\ref{5}) become 
\begin{equation}
\varphi_{0}^{2}G_{\mu\nu} - \alpha^{-2} W_{\mu\nu} = \frac{1}{2}T_{\mu\nu},
\label{10}
\end{equation}
\begin{equation}
R = 0.
\label{11}
\end{equation} 
In addition, for $\varphi = \varphi_{0}$, the MCG line element 
$ds^2 = \left(\varphi/\varphi_{0}\right)^{2}g_{\mu\nu}dx^{\mu}dx^{\nu}$ 
reduces to  
\begin{equation}
ds^2 = g_{\mu\nu}dx^{\mu}dx^{\nu}.
\label{12}
\end{equation}
The full cosmological content of MCG can be obtained from 
(\ref{10})-(\ref{12}) without loss of generality.

In order to find the MCG matter energy-momentum tensor, we consider 
the conformally invariant matter Lagrangian density \cite{Man}
\begin{equation}
\mathcal{L}_{m} = -\sqrt{-g}\bigg[S^{2}R + 6\partial^{\mu}S\partial_{\mu}S 
+ \lambda S^{4}  + \frac{i}{2}\left(\, \overline{\psi}
\gamma^{\mu}D_{\mu}\psi - D_{\mu}\overline{\psi}\gamma^{\mu}\psi \right) 
+ \mu S\overline{\psi}\psi\bigg],
\label{13}
\end{equation}
where $S$ is a scalar Higgs field, $\lambda$ and $\mu$ are 
dimensionless coupling constants, $\overline{\psi} = \psi^{\dagger}
\gamma^{0}$ is the adjoint fermion field, $D_{\mu} = \partial_{\mu} 
+ [\gamma^{\nu},\partial_{\mu}\gamma_{\nu}]/8 - [\gamma^{\nu},\gamma_{\lambda}]
\Gamma^{\lambda}\,\!\!_{\mu\nu}/8$ ($\Gamma^{\lambda}\,\!\!_{\mu\nu}$ is the 
Levi-Civita connection), and $\gamma^{\mu}$ 
are the general relativistic Dirac matrices, which satisfy the anticommutation 
relation $\{\gamma^{\mu},\gamma^{\nu}\} = 2g^{\mu\nu}$. 
The variation of (\ref{13}) with respect to $S$, $\overline{\psi}$ 
and $\psi$ gives the field equations
\begin{equation}
\left(12\nabla^{\mu}\nabla_{\mu} - 2R \right) S 
- 4\lambda S^3 - \mu \overline{\psi}\psi = 0,
\label{14}
\end{equation}
\begin{equation}
i\gamma^{\mu}D_{\mu}\psi + \mu S \psi = 0,
\label{15}
\end{equation}
\begin{equation}
iD_{\mu}\overline{\psi}\gamma^{\mu} - \mu S \overline{\psi} = 0.
\label{16}
\end{equation}
Substituting (\ref{13}) into (\ref{9}), and using (\ref{14})-(\ref{16}), 
we obtain the energy-momentum tensor
\begin{eqnarray}
T_{\mu\nu} &=& 2g_{\mu\nu}\nabla^{\rho}S\nabla_{\rho}S
-8\nabla_{\mu}S\nabla_{\nu}S +4 S\nabla_{\mu}\nabla_{\nu} S  
- g_{\mu\nu}S\nabla^{\rho}\nabla_{\rho} S \nonumber \\ &&
+ 2S^{2}\left(R_{\mu\nu} - \frac{1}{4}g_{\mu\nu}R\right) 
+ T^{f}_{\mu\nu},
\label{17}
\end{eqnarray}
where
\begin{equation}
T^{f}_{\mu\nu} = \frac{i}{4}\big(\, \overline{\psi}
\gamma_{\mu}D_{\nu}\psi - D_{\nu}\overline{\psi}\gamma_{\mu}\psi 
+ \overline{\psi}\gamma_{\nu}D_{\mu}\psi - D_{\mu}\overline{\psi}\gamma_{\nu}
\psi \big) + \frac{1}{4}g_{\mu\nu}\mu S\overline{\psi}\psi
\label{18}
\end{equation}
is the fermion energy-momentum tensor.

Considering that, at scales below the electroweak scale, the Higgs field 
acquires a spontaneously broken constant vacumm expectation value $S_{0}$, 
and taking an incoherent average of $T^{f}_{\mu\nu}$ over all the fermionic modes 
propagating in a Robertson-Walker background, we find that (\ref{17}) becomes the 
energy-momentum tensor of a dynamical perfect fluid 
\begin{equation}
T_{\mu\nu} = 2 S_{0}^{2}\left(R_{\mu\nu} - \frac{1}{4}g_{\mu\nu}R\right)
+ \left( \rho + \frac{p}{c^2} \right)u_{\mu}u_{\nu} + g_{\mu\nu}p 
- g_{\mu\nu}c^2\rho_{\Lambda},
\label{19}
\end{equation}
where $\rho$ is the mass density of the usual kinematic perfect fluid, 
$p$ is the pressure of the kinematic perfect fluid, $u^{\mu}$ is the four-velocity 
of the kinematic perfect fluid, which is normalized to $u^{\mu}u_{\mu} = - c^2$, and 
$c^2\rho_{\Lambda}$ is the vacuum energy (dark energy) density.

Taking the trace of (\ref{19}) and substituting into the trace of (\ref{10}), 
we find
\begin{equation}
-\varphi_{0}^{2}R = \frac{1}{2}\left( - c^{2}\rho 
+ 3p - 4c^{2}\rho_{\Lambda}  \right).
\label{20}
\end{equation}
The additional use of (\ref{11}) then gives the relation\footnote{It is worth 
noting that $\rho_{\Lambda}$ carries the fermion four-momentum and thus it 
evolves over time. This means that the relation (\ref{21}) does not limit the matter 
content of the MCG universe as it would be in the case of a constant $\rho_{\Lambda}$.}
\begin{equation}
c^{2}\rho_{\Lambda} = \frac{1}{4}(3p - c^{2}\rho).
\label{21}
\end{equation}
Substituting this relation back into (\ref{19}), we obtain
\begin{equation}
T_{\mu\nu} = 2S_{0}^{2}\left(R_{\mu\nu} - \frac{1}{4}g_{\mu\nu}R\right) 
+ \left( \rho + \frac{p}{c^2} 
\right)u_{\mu}u_{\nu} + \frac{1}{4}g_{\mu\nu}\left(c^{2}\rho + p \right).
\label{22}
\end{equation}
According to (\ref{22}) the vacuum energy density does not contribute 
to the dynamics of the MCG universe. This makes the theory free from the 
cosmological constant problem.


\section{Cosmological solutions}
\label{sec3}


By substituting (\ref{22}) into (\ref{10}), and considering (\ref{11}), we find
\begin{equation}
2\left(\varphi_{0}^{2}-S_{0}^{2}\right)R_{\mu\nu} 
- 2\alpha^{-2}W_{\mu\nu} = \left( \rho + \frac{p}{c^2} 
\right)u_{\mu}u_{\nu} + \frac{1}{4}g_{\mu\nu}\left(c^{2}\rho + p \right).
\label{23}
\end{equation}
Then, using the Friedmann–Lema\^itre–Robertson–Walker 
(FLRW) line element
\begin{equation}
ds^{2} = - c^2dt^{2} + a(t)^2\left( \frac{dr^{2}}{1-Kr^{2}} +r^{2}d\theta^{2} 
+ r^{2}\sin^{2}\theta d\phi^{2} \right),
\label{24}
\end{equation}
and the fluid four-velocity $u^{\mu} = (c, 0, 0 ,0)$, 
the $00$ and $11$ components of (\ref{23}) gives
\begin{equation}
\frac{\ddot{a}}{a} = -\frac{8\pi G_{\mathrm{eff}}}{3c^2}\left(c^{2}\rho + p \right),
\label{25}
\end{equation}
\begin{equation}
\frac{\ddot{a}}{a} + 2\left( \frac{\dot{a}}{a}\right)^2 + 2 \frac{K}{a^2}
=   \frac{8\pi G_{\mathrm{eff}}}{3c^2}\left(c^{2}\rho + p \right),
\label{26}
\end{equation}
where the dot denotes $d/dt$, $a = a(t)$ is the scale factor, 
$K=$ -1, 0 or 1 is the spatial curvature, and
\begin{equation}
G_{\mathrm{eff}} = \frac{3c^2}{64\pi\left(\varphi_{0}^{2}-S_{0}^{2}\right)}
\label{27}
\end{equation}
is an effective gravitational constant.

Subtracting (\ref{25}) from (\ref{26}), we obtain 
\begin{equation}
\left(\frac{\dot{a}}{a}\right)^2 + \frac{K}{a^2}  
=  \frac{8 \pi G_{\mathrm{eff}}}{3c^2}\left(c^{2}\rho + p \right).
\label{28}
\end{equation}
The combination of this equation with (\ref{25}) gives the energy continuity 
equation 
\begin{equation}
\frac{d}{dt}\left[\left(c^{2}\rho + p \right) a^4 \right] = 0,
\label{29}
\end{equation}
from which follows that
\begin{equation}
c^{2}\rho(t) + p(t) = \left(c^{2}\rho_{0} + p_{0}\right)
\left(\frac{a_{0}}{a}\right)^{4},
\label{30}
\end{equation}
where, from now on, the subscript $0$ denotes values at the present time 
$t_{0}$. 

We can write (\ref{28}) in the usual form
\begin{equation}
\Omega + \Omega_{K} = 1, 
\label{31}
\end{equation} 
where 
\begin{equation}
\Omega = \frac{8\pi G_{\mathrm{eff}}}
{3c^{2}H^2}\left(c^{2}\rho + p \right), \ \ \ \ \ 
\Omega_{K} = - \frac{K}{a^2H^2}, 
\label{32}
\end{equation}
are dimensionless density parameters, and
\begin{equation}
H = \frac{\dot{a}}{a}
\label{33}
\end{equation}
is the Hubble constant. By using (\ref{30}) in (\ref{31}), we arrive at
\begin{equation}
\left(\frac{\dot{a}}{a_{0}H_{0}}\right)^2 
=  \Omega_{0}\left(\frac{a_{0}}{a}\right)^{2} + \Omega_{0K},
\label{34}
\end{equation}
where
\begin{equation}
\Omega_{0} = \frac{8\pi G_{\mathrm{eff}}}
{3c^{2}H_{0}^{2}}\left(c^{2}\rho_{0} + p_{0} \right), \ \ \ \ \ 
\Omega_{0K} = - \frac{K}{a_{0}^{2}H_{0}^{2}}. 
\label{35}
\end{equation}

The combination of (\ref{31}) and (\ref{34}) gives 
\begin{equation}
dt = \frac{dx}{H_{0}(1-\Omega_{0} + \Omega_{0} x^{-2})^{1/2}},
\label{36}
\end{equation}
where $x = a/a_{0}$. It follows from (\ref{36}) that the time at which light 
emitted from a cosmological source reaches the earth with redshift $z$ is 
given by 
\begin{equation}
t = \frac{1}{H_{0}} \int^{1/(1+z)}_{0}{\frac{dx}{(1-\Omega_{0} 
+ \Omega_{0} x^{-2})^{1/2}}},
\label{37}
\end{equation}
where we considered that the zero of time corresponds to an infinite redshift, 
and
\begin{equation}
1+z = \frac{a_{0}}{a}.
\label{38}
\end{equation}

By considering the redshift equal to zero in (\ref{37}), we find the present 
age of the MCG universe
\begin{equation}
t_{0} = \left(\frac{1-\sqrt{\Omega_{0}}}{1-\Omega_{0}} \right)\frac{1}{H_{0}}.
\label{39}
\end{equation}
We can see from this equation that 
\begin{equation}
0 < t_{0} < \frac{1}{2H_{0}}   
\label{40}
\end{equation}
for a closed universe ($\Omega_{0} > 1$), 
\begin{equation}
t_{0} = \frac{1}{2H_{0}}  
\label{41}
\end{equation}
for a flat universe ($\Omega_{0} = 1$), and
\begin{equation}
\frac{1}{2H_{0}} < t_{0} < \frac{1}{H_{0}}  
\label{42}
\end{equation}
for an open universe ($\Omega_{0} < 1$), which means 
that only the open MCG universe is consistent with the available observational 
values of $H_{0}$ and $t_{0}$.

Assuming that the MCG universe is open ($K=-1$), we can write (\ref{34}) in the form 
\begin{equation}
\dot{a}^2 = \frac{a_{0}^{4}H_{0}^{2}\Omega_{0}}{a^2} + 1,
\label{43}
\end{equation}
whose solution is given by
\begin{equation}
a(t) = \left[\left(2a_{0}^{2}H_{0}\sqrt{\Omega_{0}}\right)t + t^2\right]^{1/2}.
\label{44}
\end{equation}
Substituting (\ref{44}) into the deceleration parameter
\begin{equation}
q = -\frac{\ddot{a}a}{\dot{a}^2},
\label{45}
\end{equation}
we find
\begin{equation}
q(t) = \frac{a_{0}^{4}H_{0}^{2}\Omega_{0}}{(a_{0}^{2}H_{0}\sqrt{\Omega_{0}}
+t)^{2}}.
\label{46}
\end{equation}

According to (\ref{44}) and (\ref{46}) the MCG universe begins with a big bang 
at $t=0$ and continues to expand decelerated forever, becoming flat as 
$t \rightarrow \infty$. Despite this being the same behavior of a pure radiation 
universe with negative curvature, the evolution of the MCG universe is ruled 
by $G_{\mathrm{eff}}$ in the place of $G$. Noting that the vacuum expectation 
values of $S$ and $\varphi$ vanish above the electroweak and Planck scales, 
respectively, we have that $G_{\mathrm{eff}} = 3c^2/64\pi\varphi_{0}^{2}$ between 
the electroweak and Planck scales, and $G_{\mathrm{eff}} = \infty$ above the 
Planck scale. This behavior of the effective gravitational constant plays an 
essential role in the evolution of the early MCG universe.

It is worth noting that MCG has a massive spin-$2$ field with negative 
energy, which can lead to instabilities in the classical solutions of 
the theory. The stability analysis of the MCG cosmological solution requires 
the linearization of (\ref{43}) about the perturbed scale factor
\begin{equation}
a'(t) = a(t)[1+\varepsilon(t)],
\label{47}
\end{equation}
where $a(t)$ is given by (\ref{44}) and $|\varepsilon(t)| \ll 1$. 
This linearization gives the perturbation equation
\begin{equation}
\dot{\varepsilon}\left(\dot{a}a^3\right)+\varepsilon
\left[a^2c^2+2\left(a_{0}^{2}H_{0}\sqrt{\Omega_{0}}\right)^2\right]=0.
\label{48}
\end{equation}
The solution of (\ref{48}), which is given by
\begin{equation}
\varepsilon(t) = \frac{a_{0}^{2}H_{0}\sqrt{\Omega_{0}}
+t}{\left(2a_{0}^{2}H_{0}\sqrt{\Omega_{0}}\right)t + t^2},
\label{49}
\end{equation}
do not grow unboundedly with time. This implies that the 
MCG cosmological solution is stable.


\section{Cosmological experimental tests}
\label{sec4}


In order to compare the cosmological solution of MCG with cosmological 
observations, we must find the luminosity distance
\begin{equation}
d_{L}(z) = a_{0}r(z)(1+z),
\label{50}
\end{equation}
where $r(z)$ is the radial distance of a cosmological light source that is 
observed now on earth with redshift $z$.

In an open universe such as the MCG universe, the radial distance is determined 
by the equation of the radial worldline of a light ray
\begin{equation}
\int_{0}^{r(z)}\frac{dr}{\sqrt{1+r^2}} = \int_{t(z)}^{t_{0}}\frac{c\,dt}{a(t)}.
\label{51}
\end{equation}
By using (\ref{36}) in (\ref{51}), integrating both sides, substituting the 
result into (\ref{50}), and making some algebra, we find
\begin{equation}
\frac{H_{0}d_{L}}{c} = \frac{\left(1+\sqrt{1-\Omega_{0}}\right)^{2}
\left(1+z\right)^{2} 
- \left(\sqrt{1-\Omega_{0}}+\sqrt{1+2\Omega_{0}z+\Omega_{0}z^{2}}\right)^{2}}
{2\left(\sqrt{1-\Omega_{0}}+\sqrt{1+2\Omega_{0}z+\Omega_{0}z^{2}}\right)
\left(1+\sqrt{1-\Omega_{0}}\right)\left(\sqrt{1-\Omega_{0}}\right)}. 
\label{52}
\end{equation}

In order to estimate the values of $H_{0}$ and $\Omega_{0}$, we use the Pantheon 
compilation with 1048 Type Ia supernovae (SNIa) data \cite{Scolnic}. The 
procedure consists in compare, for each SNIa at redshift $z_{i}$, the observed 
distance modulus $\mu_{\mathrm{obs}}(z_{i})$ ($\equiv m - M$, where $m$ is the 
apparent magnitude of the SNIa and $M=-19.3$ its absolute magnitude) with the 
theoretical distance modulus $\mu_{\mathrm{th}}(z)$ defined as 
\begin{equation}
\mu_{\mathrm{th}}(z) = 5\log_{10}d_{L}(z) + 25,
\label{53}
\end{equation}
where $d_{L}$ is measured in Mpc. The best-fit values of $H_{0}$ and $\Omega_{0}$ 
are determined by an iterative minimization of the function\footnote{Although 
the $\chi^{2}$ procedure has a degeneracy between the Hubble 
constant and the absolute magnitude, it is useful to estimate the current 
values of the cosmological parameters with a reasonable confidence 
level. In order to find more accurate estimates, we must use model-independent 
procedures such as the Bayesian approach. However, due to the greater 
complexity of these procedures, we will leave them for future works.}
\begin{equation}
\chi^{2}(\Omega_{0},H_{0}) = \sum_{i=1}^{1048}\frac{[\mu_{\mathrm{obs}}(z_{i})  
- \mu_{\mathrm{th}}(z_{i})]^2}{\sigma_{i}^{2}},
\label{54}
\end{equation}
where $\sigma_{i}$ is the uncertainty on $\mu_{\mathrm{obs}}(z_{i})$.

\begin{figure}[h!b]
	\centering \includegraphics[scale=0.4]{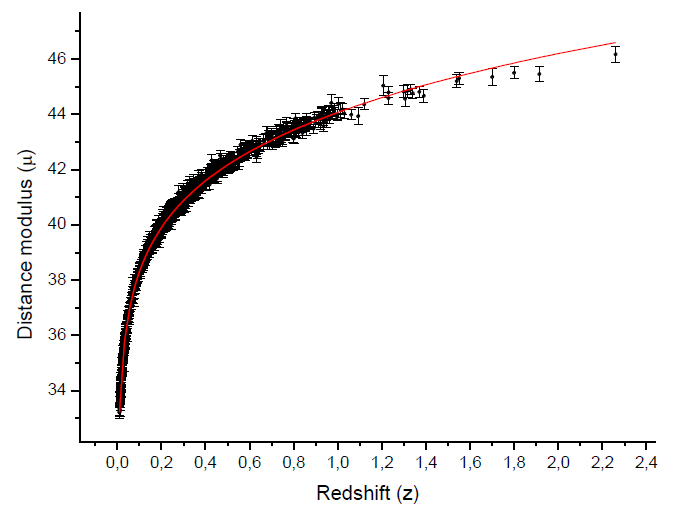}
	\caption{\label{f1} Hubble diagram for the Pantheon compilation with 1048 SNIa data. 
	The red line represents the best-fitted MCG model.}
\end{figure}

The Hubble diagram with the 1048 SNIa from the Pantheon compilation and the 
best-fit MCG model is shown in FIG. \ref{f1}. We find that the best-fit 
values of the MCG parameters are given by 
\begin{equation}
H_{0} = 69.05 \pm 0.14 \,\,  \mathrm{km} \, \mathrm{s}^{-1} 
\mathrm{Mpc}^{-1},
\label{55}
\end{equation}
\begin{equation}
\Omega_{0} =  10^{-17},
\label{56}
\end{equation}
with $\chi^{2}_{\mathrm{min}}/\mathrm{dof} = 0.99$, where dof is the logogram 
of degree of freedom. 

We can consider that the current MCG universe is dominated by baryonic 
and dark matters with combined mass densities 
$\rho_{0} \approx 2.67 \times 10^{-27}\,\mathrm{Kg\,m^{-3}}$ and 
pressures $p_{0}\approx 0$. Using these values, (\ref{55}), and (\ref{56}) in 
(\ref{35}),  we find the effective gravitational constant 
\begin{equation}
G_{\mathrm{eff}} \approx 2.23 \times 10^{-27} 
\,\,\mathrm{m^{3}\,kg^{-1}\,s^{-2}},
\label{57}
\end{equation}
and the current scale factor
\begin{equation}
a_{0} \approx 4.47 \times 10^{17} \,\,\mathrm{s}.
\label{58}
\end{equation}
Finally, the substitution of (\ref{55}) and (\ref{56}) into (\ref{39}) gives 
the current age of the MCG universe
\begin{equation}
t_{0} = 14.19 \pm 0.03 \,\, \mathrm{Gyr},
\label{59}
\end{equation}
which is consistent with the $14$ Gyr estimated from old globular clusters 
\cite{Pont}.  Further analysis is needed to see if the MCG universe 
accommodates the age of old quasars such as the APM 08279+5255\footnote{The age 
determination of old high redshift objects such as APM 08279+5255 
requires a rigorous statistical analysis involving many astrophysical 
constraints, which is beyond the scope of this paper.}.

It is worth noting that despite the extremely low values in (\ref{56}) 
and (\ref{57}), MCG fits to the SNIa data as well as the $\Lambda$CDM 
model. To see if MCG also fits well to overlapping SNIa, cosmic microwave 
background (CMB) and baryon acoustic oscillations (BAO)
data, it is necessary the development of a theory for the growth of 
inhomogeneities in the model, which would be different from the one used in 
the $\Lambda$CDM model due to the contribution of the Bach tensor. Since this 
theory has not yet been developed, this is another topic that we will leave 
for future works.


\section{Final remarks}
\label{sec5}


We have shown in this paper that the negative curvature cosmological solution 
of MCG is compatible with SNIa data without presenting the 
cosmological constant problem. The early MCG cosmology, in 
particular CMB production and nucleosynthesis, and the compatibility of the 
current age of the MCG universe with the age of old high redshift objects will 
be investigated in the future.


\end{document}